\documentclass[twocolumn,showpacs,preprintnumbers,amsmath,amssymb,nofootinbib]{revtex4}

\usepackage{graphicx}
\usepackage{epsfig}
\usepackage{color}

\newcommand{\beq}{\begin{equation}}
\newcommand{\eeq}{\end{equation}}
\newcommand{\bea}{\begin{eqnarray}}
\newcommand{\eea}{\end{eqnarray}}
\newcommand{\bwd}{\begin{widetext}}
\newcommand{\ewd}{\end{widetext}}

\begin{document}
\title{Generation of Multi-Color Attosecond X-Ray Radiation Through Modulation Compression}

\author{Ji Qiang}
\email{jqiang@lbl.gov}
\affiliation{Lawrence Berkeley National Laboratory, Berkeley, CA 94720, USA}

\author{Juhao Wu}
\affiliation{SLAC National Accelerator Laboratory, Menlo Park, CA 94025, USA}

\date{\today}

\begin{abstract}
In this paper, we propose a scheme to generate tunable multi-color attosecond 
coherent X-ray radiation for future light source applications.
This scheme uses an energy chirped electron beam, a laser modulators, a laser
chirper and
two bunch compressors to generate a multi-spike prebunched kilo-Ampere
current electron beam from a few tens Ampere electron beam out of a linac.
Such an electron beam transports through a series of undulator radiators and
bunch compressors to generate multi-color coherent X-ray radiation.  
As an illustration, we present an example to generate two attosecond
pulses with $2.2$ nm and $3$ nm 
coherent X-ray radiation wavelength and more than $200$ MW peak power
using a $30$ Ampere $200$ nm laser seeded electron beam.

\end{abstract}

\maketitle


Attosecond coherent X-ray source provides an important
tool to study ultra-fast dynamic process in 
biology, chemistry, physics, and material science.
In recent years, there is growing interest in generating
attosecond X-ray radiation pulse using Free Electron Lasers
(FELs)~\cite{zholents1,zholents2,saldin2,wu2,zholents3,ding,xiang2,
zholents4}. Most of those schemes generates
a single color attosecond pulse X-ray radiation except 
that in reference~\cite{zholents4},
where two attosecond
X-ray radiation pulses with different radiation wavelength (colors)
were produced based on a recently proposed echo scheme~\cite{echo}. 
Meanwhile, multi-color attosecond X-ray radiation has 
important applications in time-resolved
experiments such as
multidimensional X-ray spectroscopy
by either exciting or probing different types of atom in a system~\cite{multcor}.
In reference~\cite{zholents4}, the 
second color attosecond pulse radiation was produced
using a few-cycle laser modulator, a bunch compressor, and
a short undulator radiator.
In this paper, we propose a scheme to generate such tunable multi-color attosecond 
coherent X-ray radiation based on an
improved modulation compression method~\cite{qiang2}.
This scheme avoids the usage of an extra laser modulator to produce 
an extra color attosecond X-radiation in the reference 8.
It also uses a low current (on the order of ten Amperes) electron beam out of a linac instead of a kilo-Ampere electron beam used in the reference 8.

\begin{figure}[tb]
   \centering
   \includegraphics*[angle=0,width=90mm]{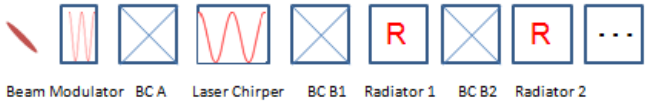}
   \caption{A schematic plot of the lattice layout of the modulation compression scheme.}
   \label{figxban3}
\end{figure}

A schematic plot of the scheme to generate multi-color attosecond X-ray
radiation is given in Fig.~\ref{figxban3}.
It consists of an energy chirped electron beam, a seeding laser modulator, a bunch compressor A,
a laser chirper, a bunch compressor B1, an undulator radiator one, a
bunch compressor B2, another undulator radiator two, and a number of
 repeating bunch compressors and radiators for multi-color X-ray radiation.
Assume an initial beam longitudinal phase space distribution of the beam
as:
\begin{eqnarray}
f(z,\delta) & = & F(z,(\delta-hz)/\sigma)
\end{eqnarray}
where
$z$ is the relative longitudinal distance with respect to the reference particle,
$\delta = \Delta E/E$ is the relative energy deviation, 
$h$ is the initial beam energy chirp, and $\sigma$ is the initial
uncorrelated energy spread.
By properly choosing 
the momentum compaction
factor of bunch compressor B such that
\begin{eqnarray}
R^b_{56} = -R^a_{56}/M
\label{rr}
\end{eqnarray}
the longitudinal phase space distribution after the bunch compressor B1
can be written as:
\begin{eqnarray}
f(z,\delta) & = & F(Mz,[\delta - M \tilde{h}z- MA\sin(kMz)]/M \sigma)
\end{eqnarray}
where
\begin{eqnarray}
M & = & 1+h_b R^a_{56} 
\label{mm}
\end{eqnarray}
represents the total modulation compression factor, 
$\tilde{h}  =  h_b/C + h$, $C$ is the compression factor from the
first bunch compressor A,
$h_b$ is the energy chirp introduced by the laser chirper, 
$R_{56}^a$ is the momentum compaction factor of the first bunch compressor,
$A$ is the modulation amplitude of the seeding laser
in the unit of the relative energy, 
and $k$ is the wave number of the seeding laser.
The above distribution function represents a compressed modulation
in a chirped beam. 
In the above equations, we have also assumed a longitudinally frozen electron beam and a linear laser chirper instead of the real sinusoidal function from the laser modulator. 

The linear chirp in Eq.~\ref{mm} from a sinusoidal laser chirper modulation can be approximated as $h_b  =  A_b \  k_b$,
where $A_b$ is the amplitude of the laser modulation, $k_b$ is
the wavenumber of the laser. 
The sinusoidal form of the energy modulation
provides periodical energy chirping/unchirping across the beam. 
If the amplitude envelop of the sinusoidal energy modulation is controlled externally,
the periodical local chirping of the beam can be controlled. From Eq.~\ref{mm},
the local modulation compression factor can be controlled across the beam.
This results in a periodically separated locally modulated beam with different
modulation wavelengths. 
For a Gaussian laser beam, the energy modulation caused by the laser
chirper can be given as: 
\begin{eqnarray}
\delta & = & \delta  + A_b \sin(k_b z) \exp(-\frac{1}{2}\frac{z^2}{\sigma_b^2})
\label{engmm}
\end{eqnarray}
where $\sigma_b$ is the rms laser pulse length.
The local linear chirp resulted from such a laser chirper at different wavelength
separation will be
\begin{eqnarray}
h_b(i\lambda) & = & A_b k_b \exp(-\frac{1}{2}\frac{(i\lambda)^2}{\sigma_b^2})
\end{eqnarray}
where $i=0,\pm 1,\pm 2, \cdots$.
Using Eq.~\ref{mm}, the resultant modulation compression factor $M$ becomes
\begin{eqnarray}
M(i\lambda) & = & 1+ A_b k_b \exp(-\frac{1}{2}\frac{(i\lambda)^2}{\sigma_b^2}) R^a_{56}
\end{eqnarray}
From above equation, we see that on either side of the Gaussian laser pulse
(i.e. $i \ge 0$, or $i \le 0$), the compressed modulation wavelength
will decrease with the increase of the separation.
To achieve the final modulation compression, the $R_{56}^b$ of the bunch 
compressor Bs after the laser chirper needs to match the condition~\ref{rr}.
For the first local chirp $h_b(0)$, this can be done by using
the bunch compressor B1. After the beam passes through the first radiator
to generate first color radiation, the second bunch compressor B2
can be used to produce locally prebunched beam corresponding to the local
chirp $h_b(\lambda)$. Such a locally prebunched beam passing through
the second radiator will generate another color X-ray. 
Following the same procedure, multi-color X-ray radiation can be generated 
by using multiple of bunch compressor and radiator pairs for different
local chirp $h_b$ and modulation compression factor. 

%

\begin{figure}[htb]
   \centering
   \includegraphics*[angle=270,width=65mm]{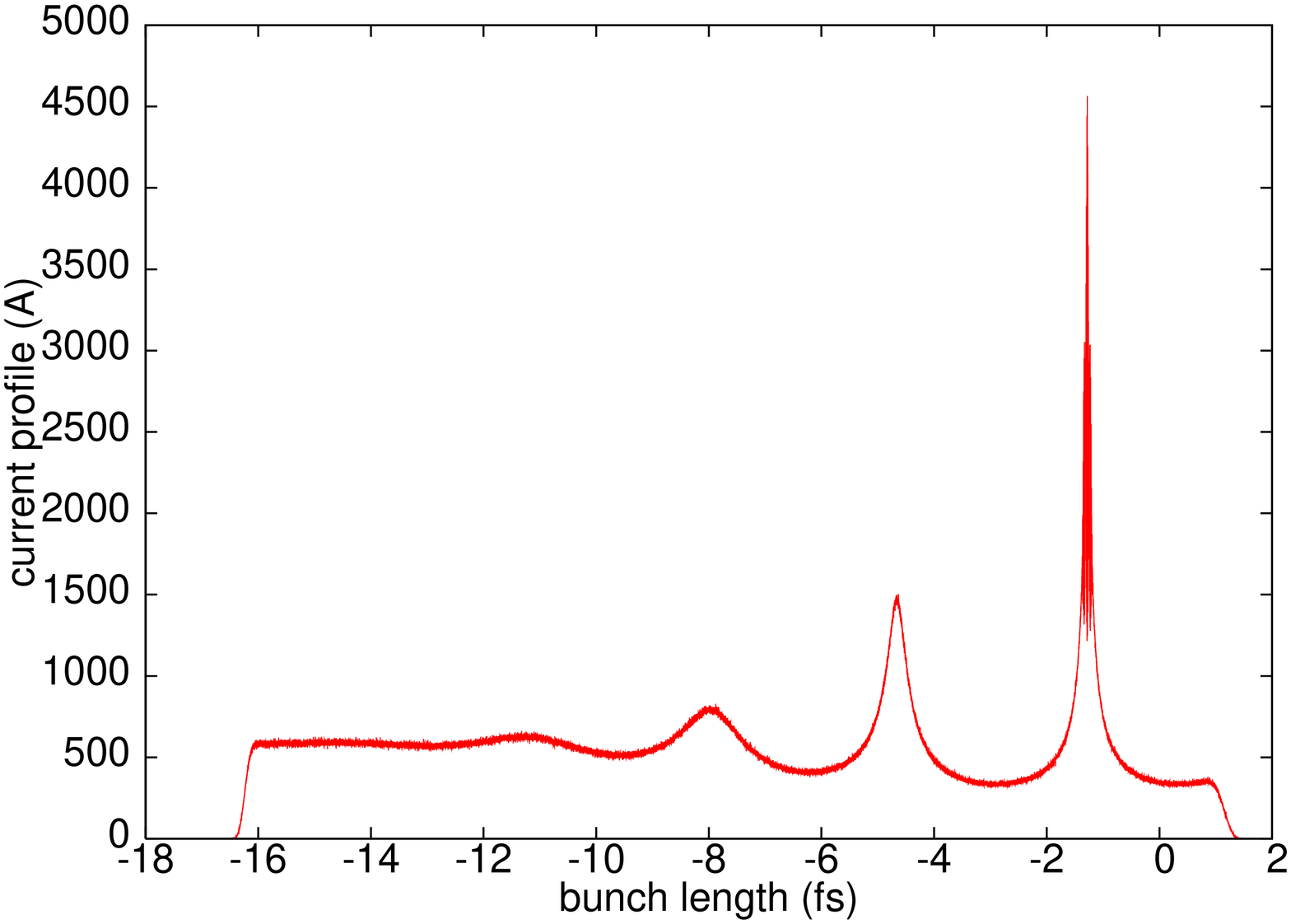}
   \includegraphics*[angle=270,width=65mm]{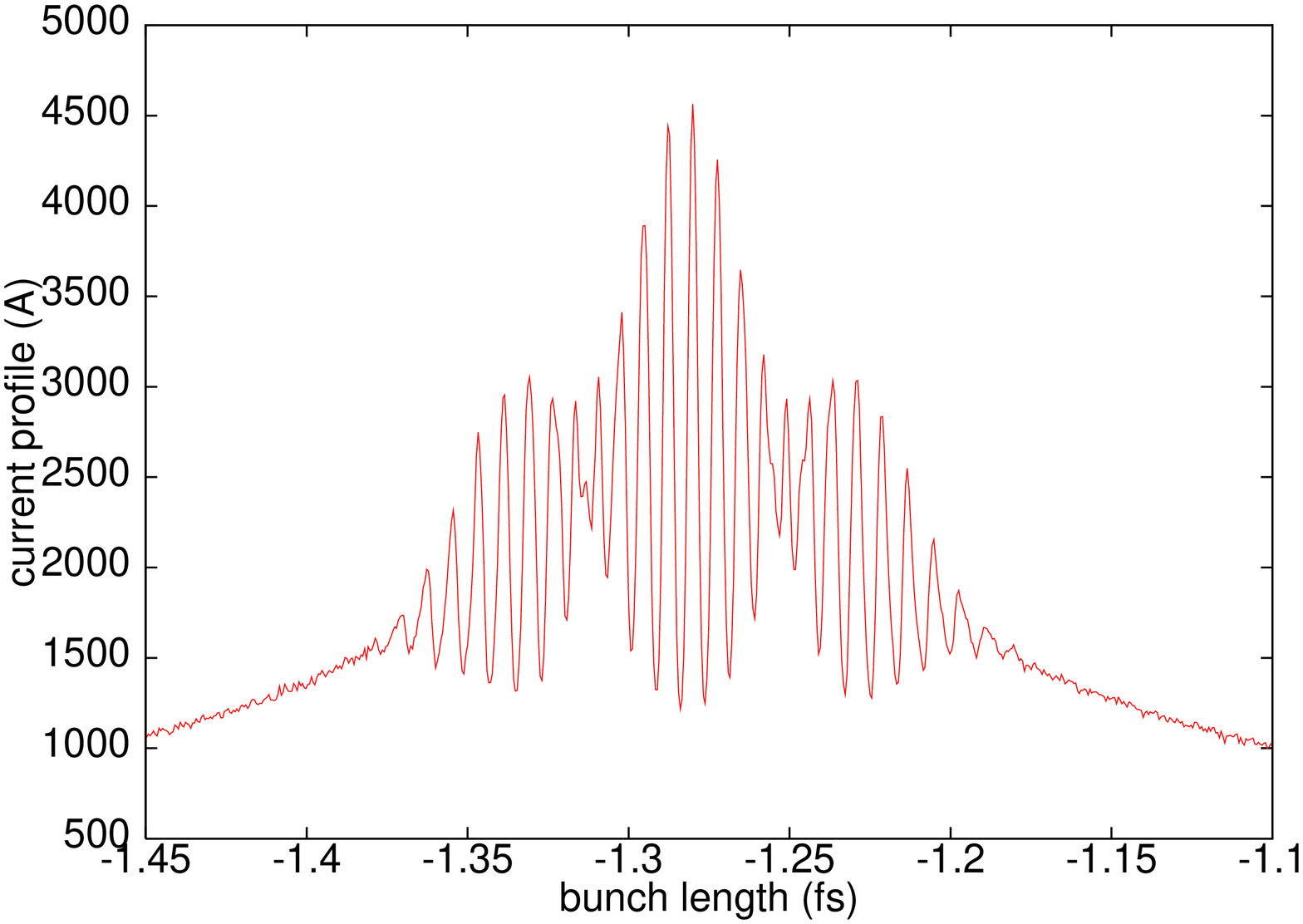}
   \caption{Beam current distribution at the end of the bunch compressor 
B1 (top) and
the zoom-in current distribution inside the first spike of the beam around $z=0$ (bottom).}
   \label{figcur}
\end{figure}
\begin{figure}[htb]
   \centering
   \includegraphics*[angle=270,width=65mm]{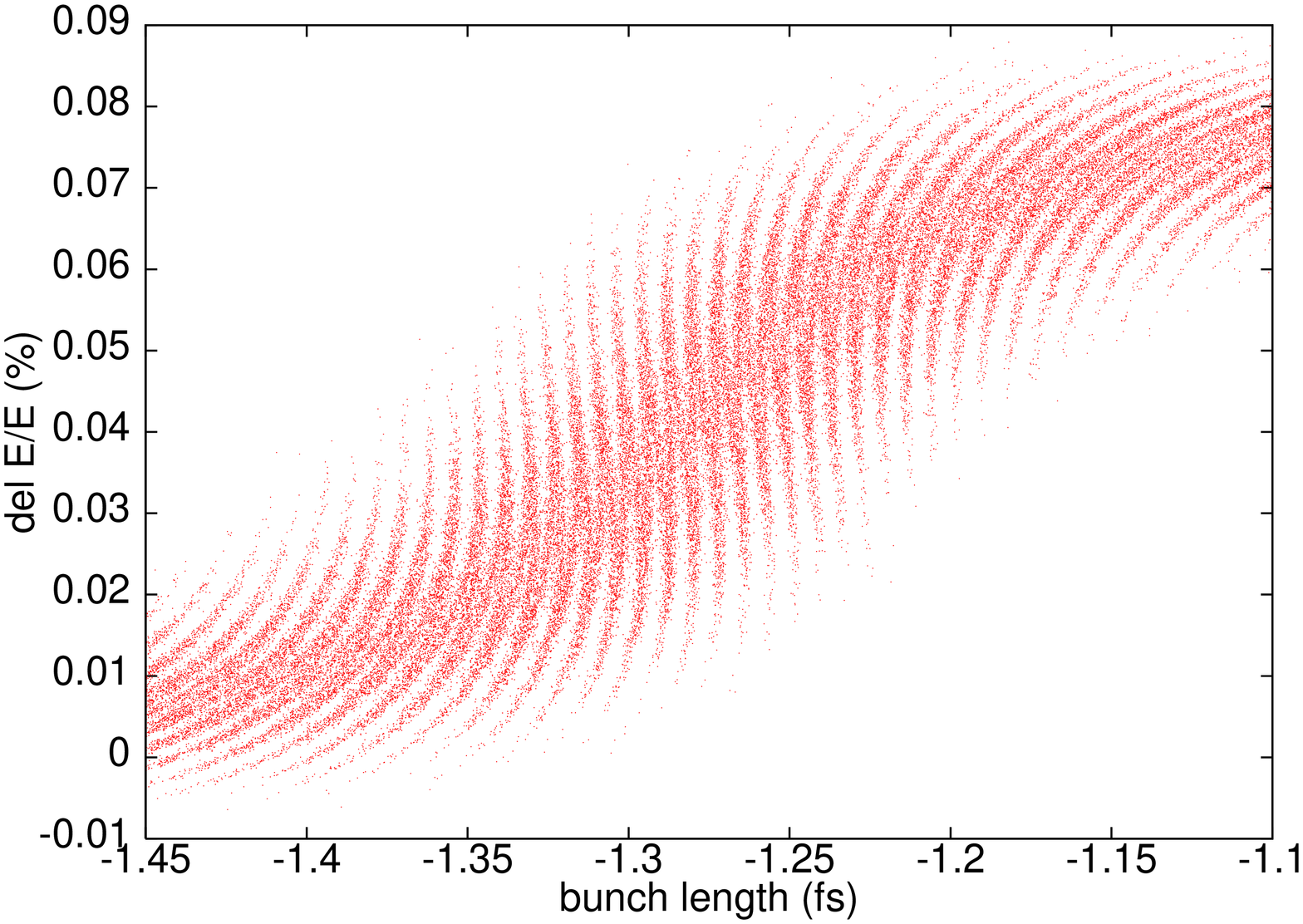}
   \includegraphics*[angle=270,width=65mm]{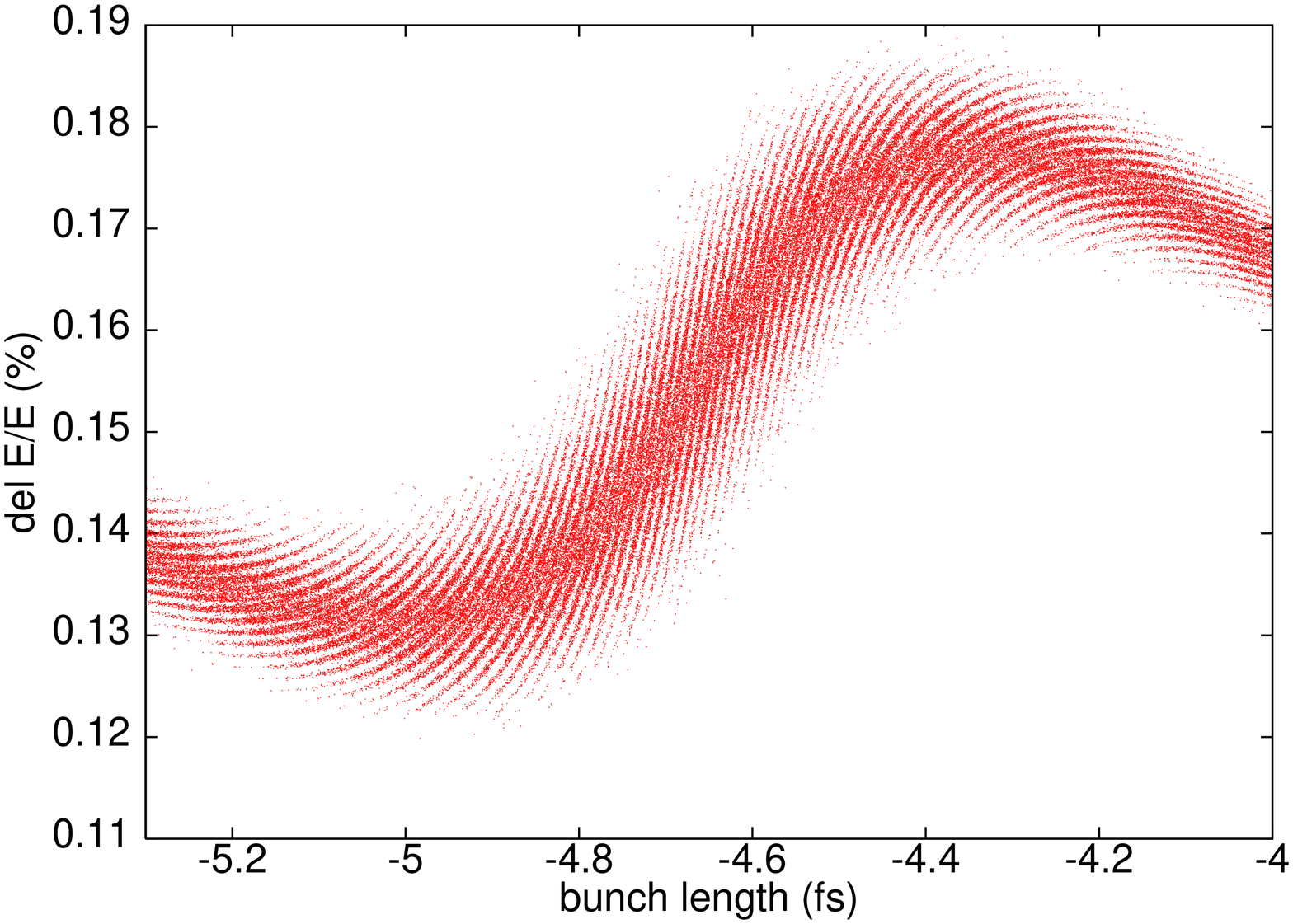}
   \caption{Longitudinal phase space inside the first spike (top) and
the second spike (bottom) of the beam at the end of the bunch
compressor B1.}
   \label{figphase}
\end{figure}
As an illustration of above scheme, we will produce two-color attosecond 
coherent X-ray radiation using a similar example in 
reference~\cite{zholents4}. 
A short uniform electron bunch ($100$ $\mu$m) with $10$ pC charge,
 $2$ GeV energy, $-54.45$ m$^{-1}$ energy-bunch length chirp, and an uncorrelated energy spread of $1 \times 10^{-6}$ is assumed at the beginning of the seeding laser modulator.
The initial normalized modulation amplitude $A$ is $1.2 \times 10^{-6}$.
Assuming $1$ Tesla magnetic field in the wiggler with a total length of $33$ cm and a period of $11$ cm, this corresponds to about
$130$ kW $200$ nm wavelength laser power.
After the modulator, we add an uncorrelated energy spread of $0.56$ keV to 
the beam to account for the synchrotron radiation effects inside the wiggler magnet.
After the initial seeding laser modulator,
the beam passes through the chicane bunch compressor A.
Here, we have assumed that the $R_{56}$ of the chicane is $1.763$ cm.
This gives a factor of $25$ compression from this bunch compressor.
Another $0.63$ keV uncorrelated energy spread is added to the beam to
account for the uncorrelated energy spread growth from the
quantum fluctuation of the
incoherent synchrotron radiation through the chicane.
After the beam passes through the chicane, 
it then transports through a short-pulse laser chirper 
with a $800$ nm resonance wavelength.
Here, we assumed that the energy modulation from the laser chirper 
follows a Gaussian envelop function as
given in Eq.~\ref{engmm} with the rms laser pulse length $\sigma_b= 1.307 \lambda_b$.
The normalized amplitude of the laser for the first local chirp  
is chosen as $6.29 \times 10^{-4}$ so that
the total modulation compression factor
is about $88.1$.
This modulation amplitude corresponds to about $38$ GW laser power using
a single wiggler period with $0.715$ T magnetic field and $22.4$ cm period length.
After the beam transports through the laser chirper, 
it passes through a dog-leg type bunch
compressor B1 that can provide opposite sign $R^b_{56}$ compared to the chicane.
For a total compression factor of $88.1$ of the first local chirp,
the $R^b_{56}$ for the bunch compressor B1 is about $-0.02$ mm.
Fig.~\ref{figcur} shows the projected current profiles at the end
of the bunch compressor B1.
Here, only half of the laser pulse is used to unchirp the
initial seeded beam in order to avoid the two locally prebunched attosecond beam
with the same modulation wavelength due to the symmetry of the laser Gaussian envelope function. 
Given an initial $30$ A beam current, 
the prebunched current inside the first spike
with about $2.2$ nm wavelength modulation reaches
about $4.5$ kA.
The width of the prebunched beam is about two hundred attoseconds.
This is set by the half wavelength of the laser chirper
and the compression of the beam.
There are other lower current
spikes besides the first spike separated by laser chirper wavelength  
inside the beam. 
Those spikes will not
contribute significantly to the $2.2$ nm attosecond radiation since
density microbunching in those spikes is very small due to
the lower modulation compression factor and the mismatch of the $R_{56}$ of
the bunch compressor B1 at those spike locations.
Fig.~\ref{figphase} shows the longitudinal phase space inside the first
spike and the second spike of the beam.
The particles inside the first spike of the beam are correctly modulated
and compressed while the particles inside the second spike are over compressed
and result in little current density modulation.

\begin{figure}[htb]
   \centering
   \includegraphics*[angle=270,width=65mm]{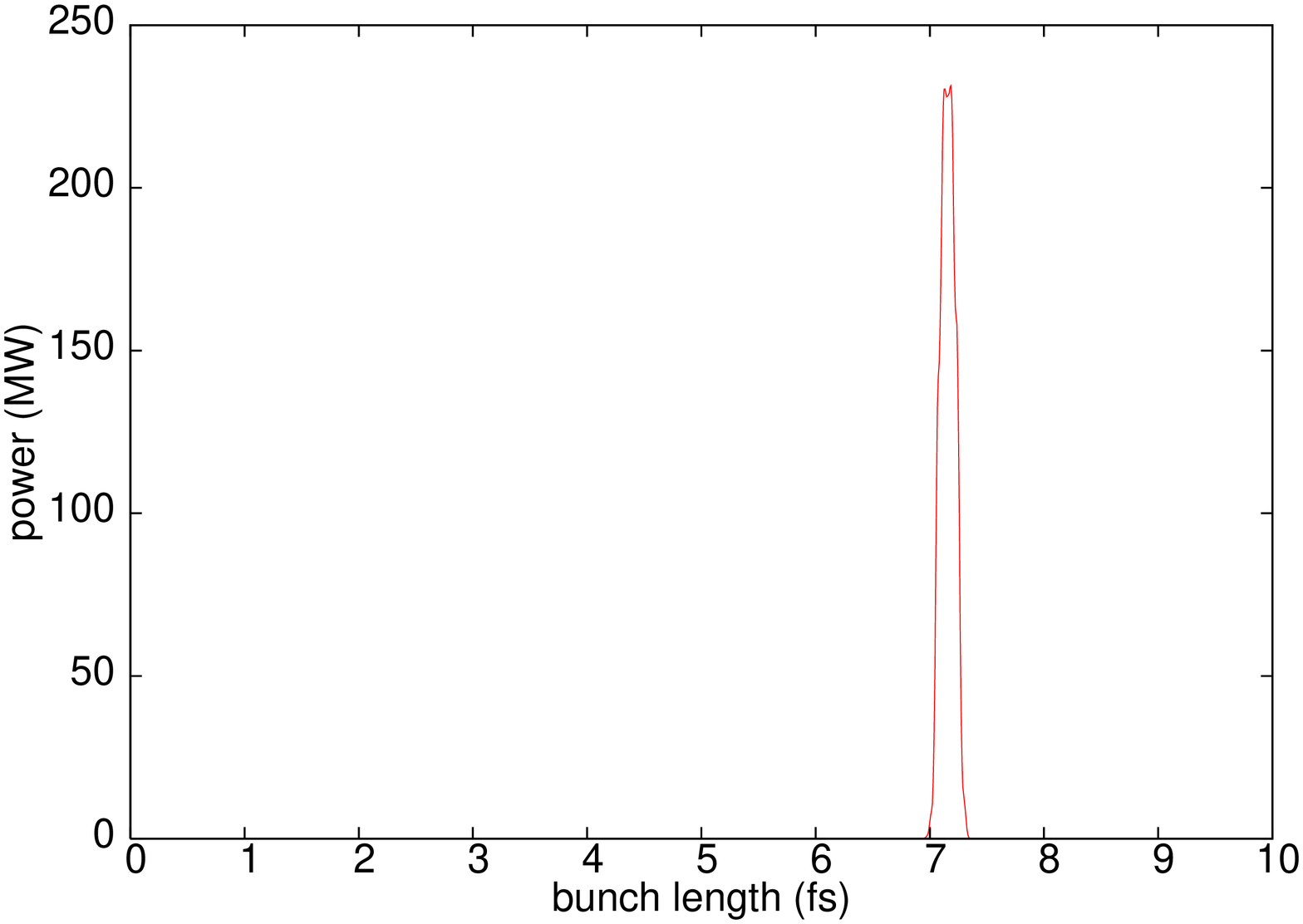}
   \includegraphics*[angle=270,width=65mm]{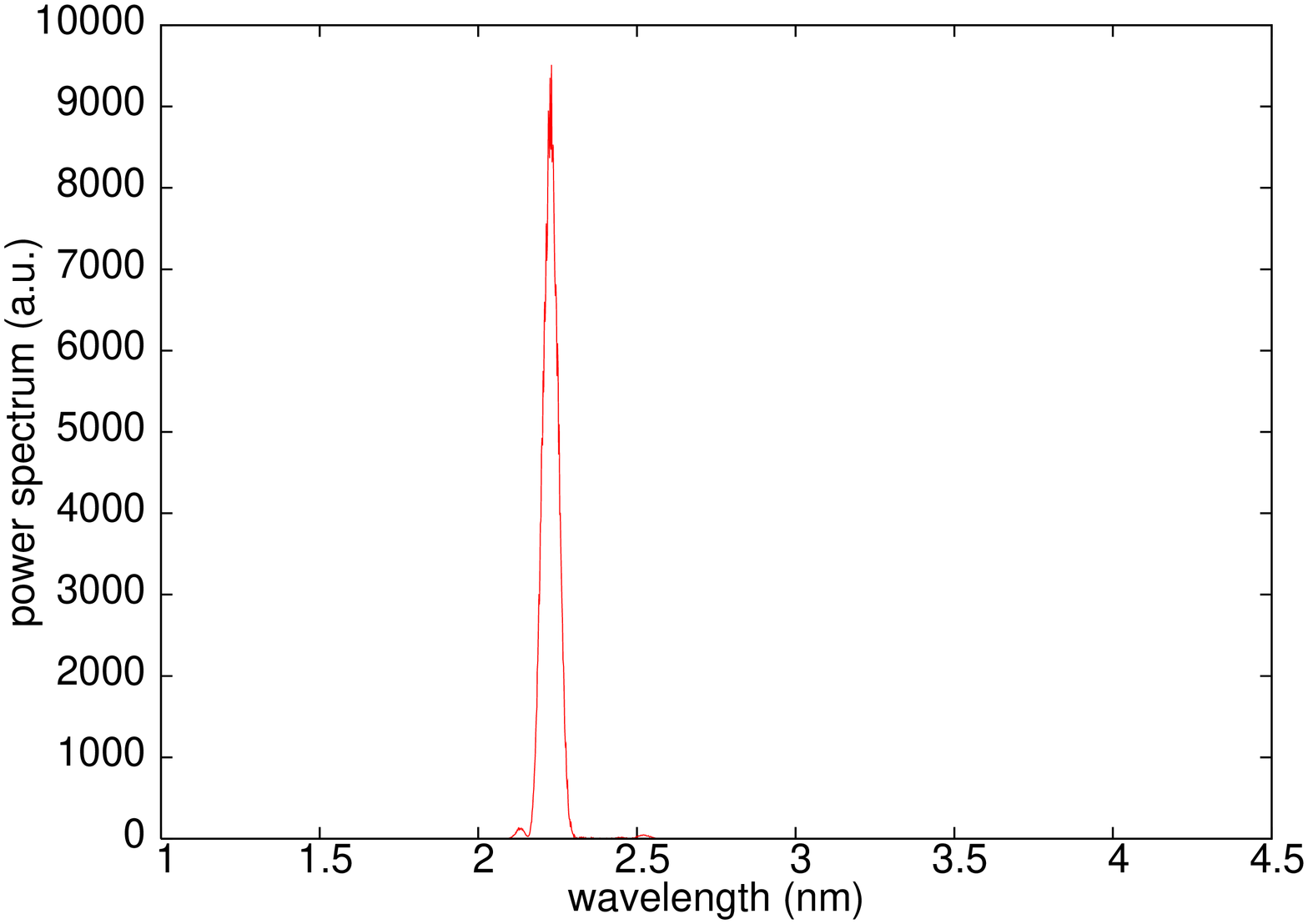}
   \caption{The radiation pulse temporal profile (top) and the radiation
pulse spectral profile (bottom) at the end of the undulator radiator 1.}
   \label{negR56sigpt1p2}
\end{figure}
The above highly prebunched beam passing through a short undulator R1 
will generate coherent attosecond X-ray 
radiation. 
Here, we have used the {\small GENESIS} simulation code~\cite{Sven99} to calculate the coherent
X-ray radiation through the short undulator radiator. The normalized emittance
of the electron beam is chosen to be $0.2$ $\mu$m.
The length of the radiator is
about $1.0$ m with an undulator period of $3.33$ cm.
The details of the radiation
properties at the end of the radiator R1 are shown in Fig.~\ref{negR56sigpt1p2}.
The full width at half maximum of the radiation pulse is about $200$ as.
The peak radiation power is beyond $220$ MW.

\begin{figure}[htb]
   \centering
   \includegraphics*[angle=270,width=65mm]{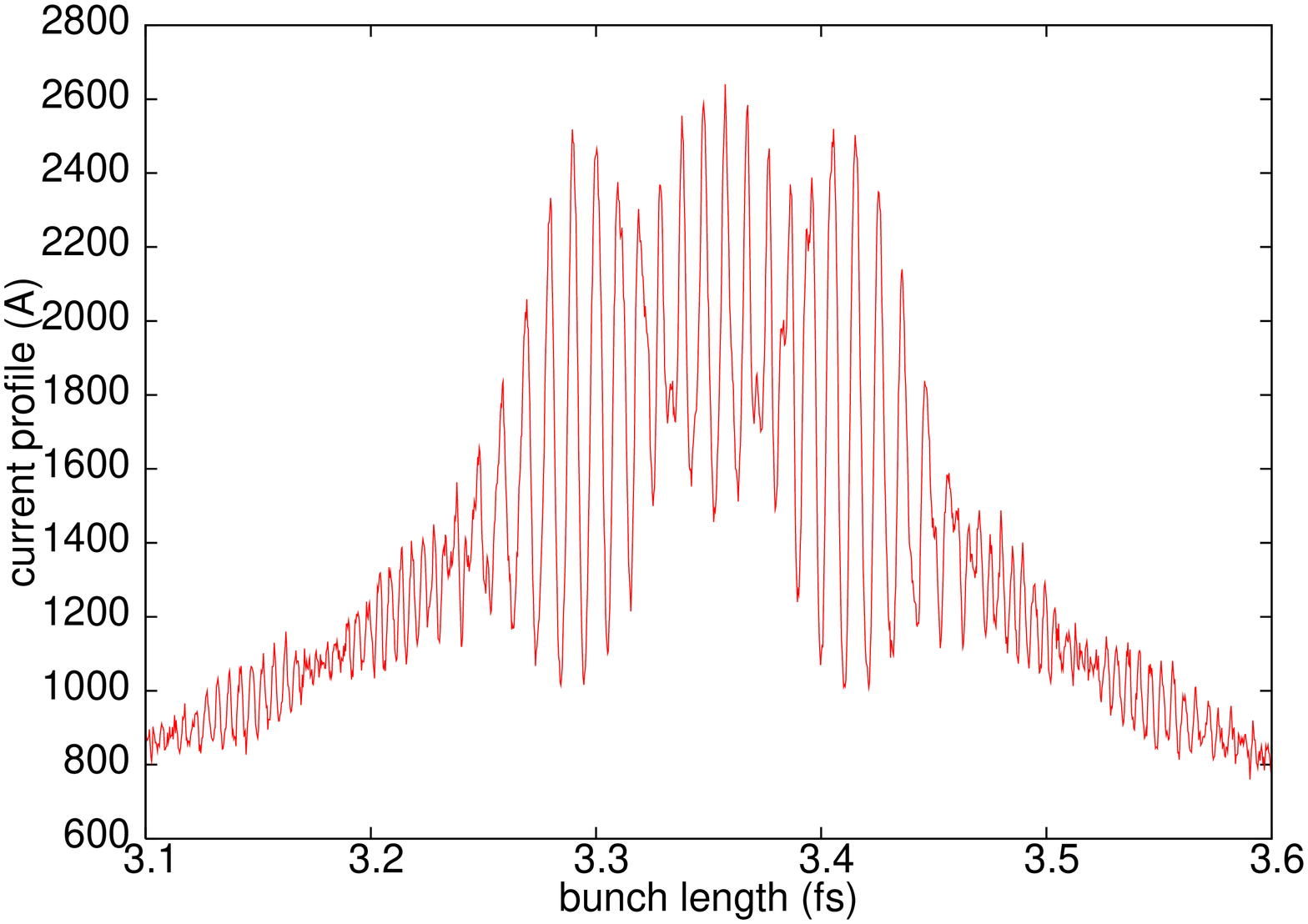}
   \includegraphics*[angle=270,width=65mm]{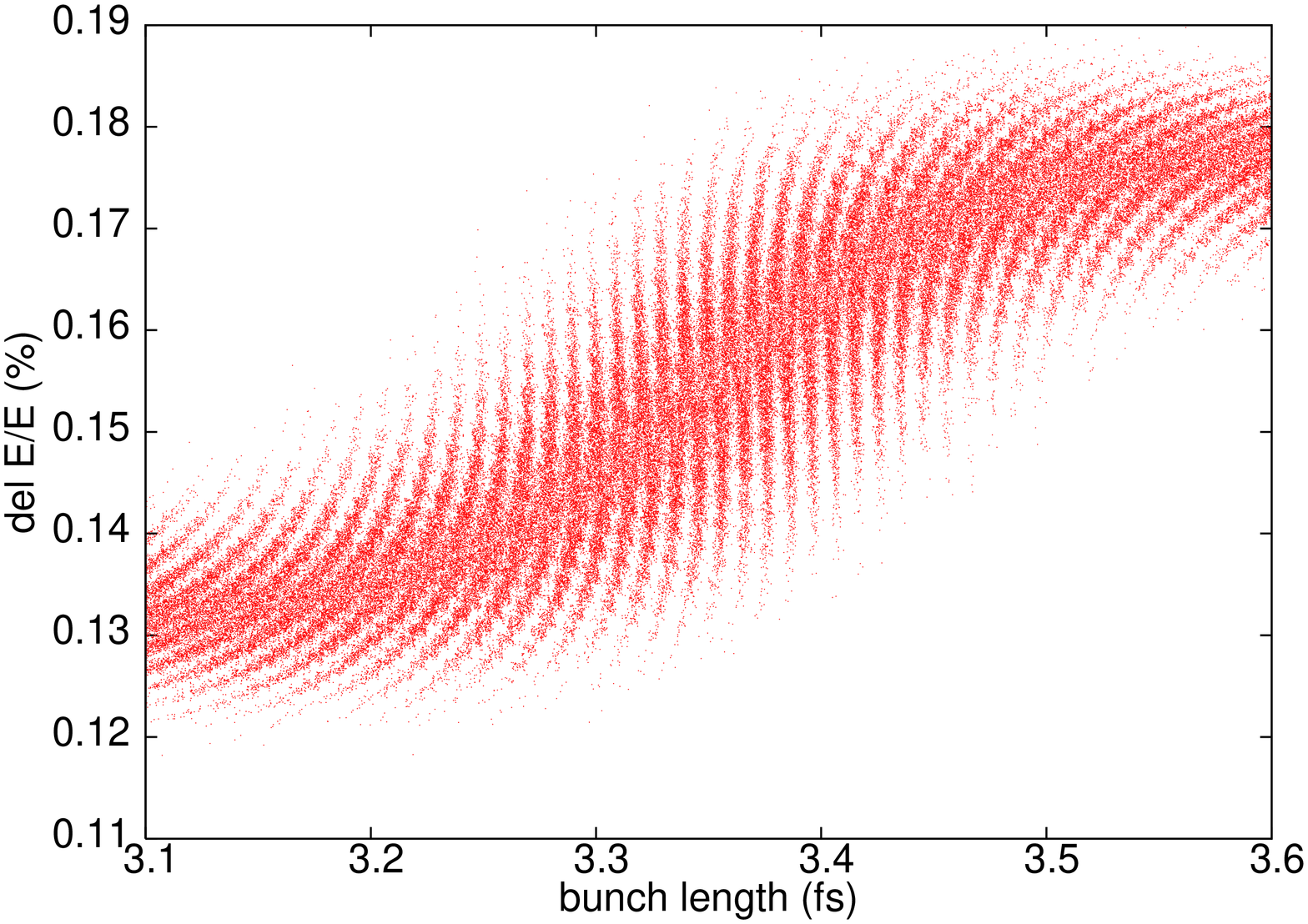}
   \caption{
Longitudinal phase space (top) and current 
profile (bottom) of the second spike at the end of the bunch compressor B2.}
   \label{beam2}
\end{figure}

\begin{figure}[htb]
   \centering
   \includegraphics*[angle=270,width=65mm]{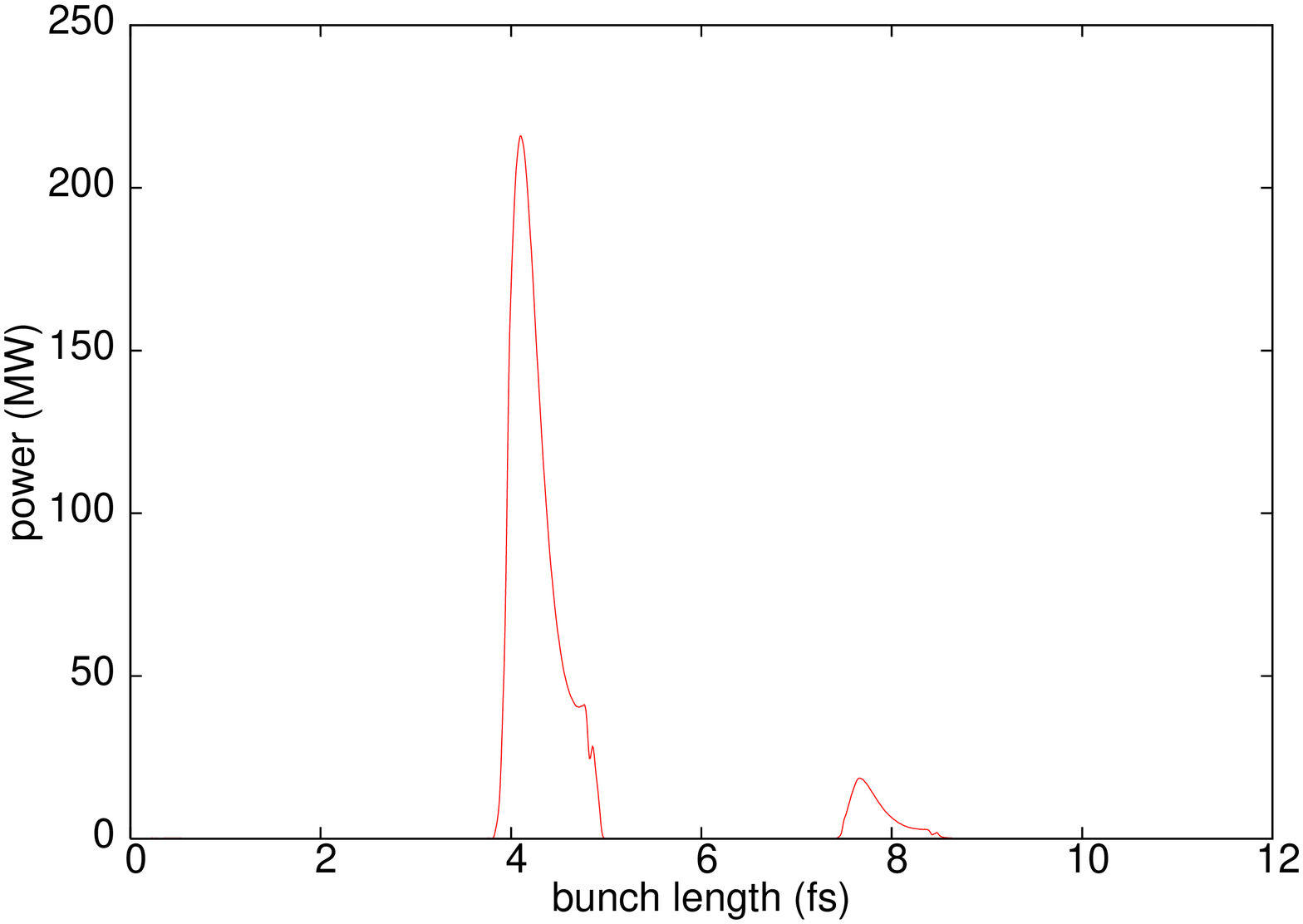}
   \includegraphics*[angle=270,width=65mm]{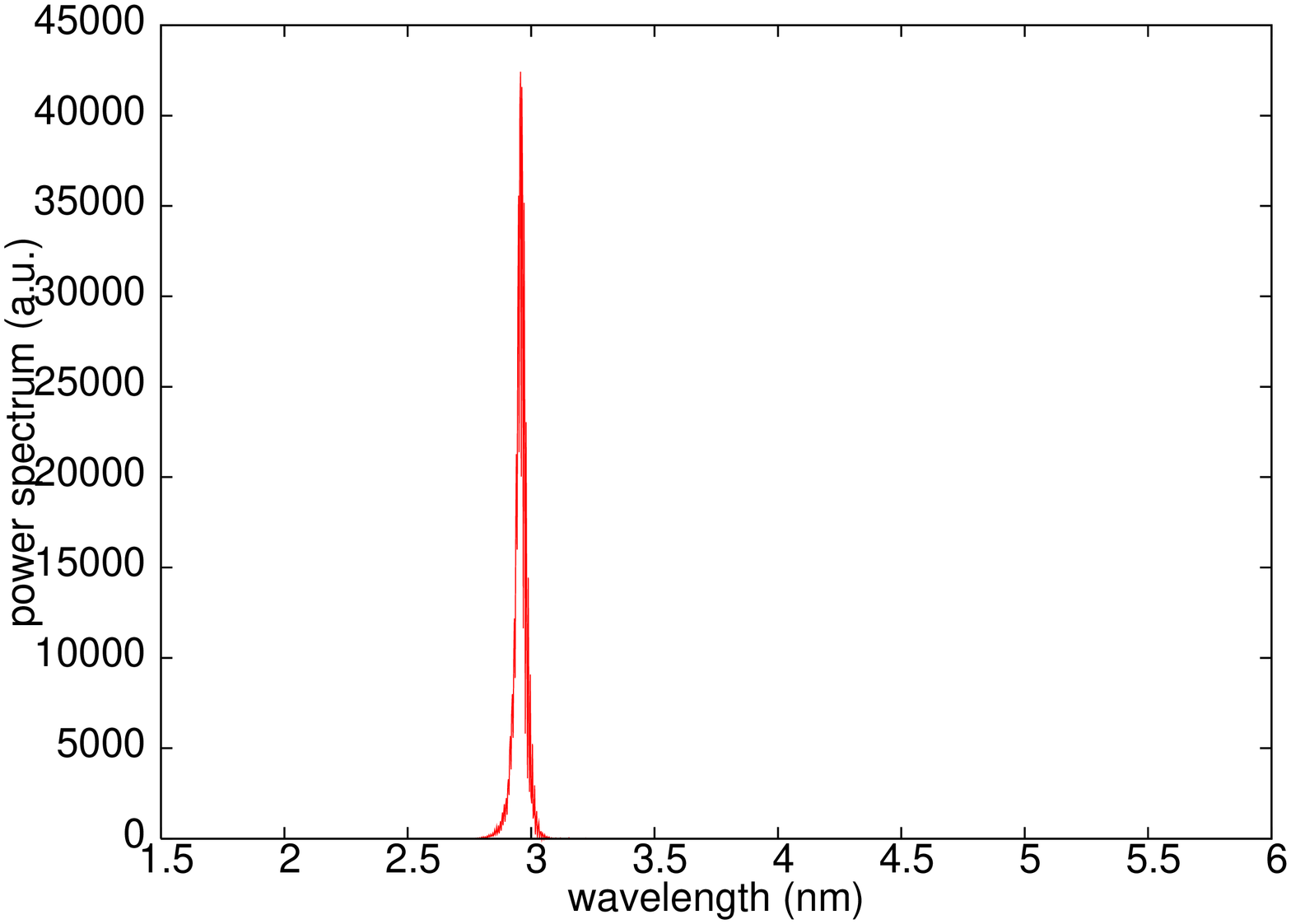}
   \caption{The radiation pulse temporal profile (top) and the radiation pulse
spectral profile (bottom) at the end of the undulator radiator 2.}
\label{rad2}
\end{figure}

After the electron beam passes through the radiator R1,
the output longitudinal particle distribution from the previous
GENESIS simulation is used to pass through the bunch compressor 
B2.
The $R_{56}$ of this
bunch compressor is chosen to be about $-61.7$ um so that the modulation
compression condition~\ref{rr} 
can be satisfied for the second current spike inside the beam.
This results in a modulation compression
factor of about $66.0$ inside that local spike. 
Fig.~\ref{beam2} shows the longitudinal phase space and current 
profile inside the second current spike of the beam at the end of 
the bunch compressor B2.
It is seen that after this bunch compressor, the beam is
significantly modulated with a wavelength about $3$ nm 
inside the second spike. 
The peak current inside this spike is about $2.6$ kA with 
a modulation width of about $200$ as.
The energy spread inside this spike is still small and there is a total upshift
of the energy due to the global energy chirp of the beam. 
Such an upshift could
also help separate the radiation of this local 
current spike from that of the other local current spikes inside the undulator. 
Fig.~\ref{rad2} shows the radiation pulse power temporal profile
and the radiation pulse spectral profile at the end of the radiator R2.
Here, we have assumed that the radiator is $4.45$ m long with an undulator
period of $4.45$ cm. 
The full width at half maximum of the radiation pulse is about $400$ as
with a radiation peak power more than $210$ MW.
The radiation pulse length
is longer than the width of the modulated current density
distribution. This is due to
the slippage of the photon pulse with respect to the
electron bunch inside the radiator.

In this paper, we proposed a scheme to generate multi-color attosecond coherent
X-ray radiation through modulation compression by using a low current
chirped electron beam.
This scheme allows one to tune the final X-ray radiation wavelength
by adjusting the compression factor. It also allows one to
control the final radiation pulse length by controlling
the laser chirper parameters and the initial compression parameters.
Some technical challenges such as keeping good synchronization between
the electron beam and the laser beam and controlling the laser power
jitter can be solved with the fast advances of the instrumentation and
the laser technology.

\begin{acknowledgments}
We would like to thank Dr. J. Corlett for useful discussions and Dr. M. Reinsch
for the discussion about the GENESIS usage.
This research used computer resources at the National Energy Research
Scientific Computing Center.
The work of JQ was supported by the U.S. Department of Energy under Contract No. DE-AC02-05CH11231 and the work of JW was supported by the U.S. Department of Energy under contract DE-AC02-76SF00515.
\end{acknowledgments}


\end{document}